# Mass Generating Antisymmetric Tensor Field Theories

Michael Kalb[*]

*Physics Department (M-AB2-02), Fairleigh Dickinson University, Madison, New Jersey 07940*
(May 28, 2009)

We study a set of theories built on a ranked sequence of antisymmetric tensor fields in $D$ dimensional space-time. These linear theories exhibit gauge invariances that are analogous to the local gauge invariance of a massless vector (Maxwell) field. By appropriate arrangements of pairs of sequentially ranked interacting fields in each member of a series of action functionals, we produce a sequence of corresponding Higgs mechanism generalizations. Therefore, mass arises in one field of each pair at the expense of the independent existence of the other. The length of the sequence is determined by $D$.

## I. INTRODUCTION

The Higgs[1] mechanism operates with local field theoretic gauge invariance to transform coupled vector and scalar fields into a massive vector field, attended by the consumption of the scalar by the vector. Subsequent work demonstrated that the mechanism could be generalized to take place between a rank-2 antisymmetric tensor field [2] (generated by the body of a relativistic string) that interacts with a vector field (produced by the end-points).[3,4] Originally both the scalar-vector and vector-tensor developments assumed a space-time dimensionality $D = 4$. For this value of $D$ a massless vector has two degrees of freedom, due to its gauge invariance, while an antisymmetric tensor has only one, due to its corresponding gauge invariance.[5] Interestingly, the set of generalized gauge invariance of the antisymmetric tensor contains the vector gauge transformations as a subset (nesting).[6]

Evidently, tensor field properties vary with $D$. In particular, easing the $D = 4$ limitation to allow higher dimensions, produces higher rank fields and allows increased field degrees of freedom for a given rank. Furthermore, since formal gauge properties continue, including nesting, the Higgs mechanism consistently and formally generalizes as well.

These new results are significant because they testify to the possibility of a set of tensor Higgs mechanisms in nature. This possibility suggests the experimental search for the Higgs boson[7] could be a broader one. Furthermore, since the possible existence and properties of these higher rank tensor fields depends explicitly on the number of space-time dimensions, their discovery (or not) may also help in understanding the roll of compactified higher dimensions in the physics.

## II. TENSOR HIGGS MECHANISM

The usual Higgs mechanism in $D$ space-time dimensions proceeds through a field theoretic action such as[8]

$$S_{(0,1)} = \kappa_D \int d^D x \left( -\tfrac{1}{4} F^{\mu\nu} F_{\mu\nu} + \tfrac{1}{2} \partial^\mu A \partial_\mu A + m_1 A_\nu \partial^\nu A + \tfrac{1}{2} m_1^2 A^\mu A_\mu \right). \tag{2.1}$$

---

[*] Electronic address: mkalb@fdu.edu

Here a vector field, $A^\mu$, interacts with a scalar field, $A$, which ultimately is consumed by the vector. The vector then acquires a mass $m_1$.[9,10] The argument begins by noting that the action displays a joint local gauge invariance under infinitesimal transformations given by

$$\delta A_\mu = \partial_\mu \lambda,$$
$$\delta A = m_1 \lambda + c. \tag{2.2}$$

Here $\lambda$ is the local gauge parameter, and $c$ is a constant representing the global part of the transformation. Next we proceed by noting that the last three terms of the action can be rewritten as $\frac{1}{2} m_1^2 B^\mu B_\mu$.

Where

$$B^\mu \equiv A^\mu + \frac{1}{m_1} \partial^\mu A. \tag{2.3}$$

Finally, the action becomes

$$S_{(0,1)} = \kappa_D \int d^D x \left( -\tfrac{1}{4} G^{\mu\nu} G_{\mu\nu} + \tfrac{1}{2} m_1^2 B^\mu B_\mu \right). \tag{2.4}$$

Where

$$G^{\mu\nu} \equiv \partial^\mu B^\nu - \partial^\nu B^\mu = F^{\mu\nu}. \tag{2.5}$$

Extremizing this action leads to a Klein-Gordon equation for $B^\mu$ with mass $m_1$.

Our generalized Higgs discussion begins by briefly examining the action for a free massless vector (Maxwell) potential field in $D$ space-time dimensions.

$$S_{(1)} = \kappa_D \int d^D x \left( -\tfrac{1}{4} F^{\mu\nu} F_{\mu\nu} \right). \tag{2.6}$$

Here

$$F^{\mu\nu} \equiv \partial^\mu A^\nu - \partial^\nu A^\mu \tag{2.7}$$

is the antisymmetric tensor field strength obtained from the vector potential $A^\mu$. Finding an extremum of the action under the variation $\delta A_\nu$ leads to the Euler-Lagrange equations,

$$\partial_\mu F^{\mu\nu} = 0. \tag{2.8}$$

The highlight of this field theory is the local gauge invariance of the action (2.6). Thus, if [11]

$$\delta A^\mu = \partial^\mu \lambda, \tag{2.9}$$

then

$$\delta F^{\mu\nu} = 0, \tag{2.10}$$

and

$$\delta S_{(1)} = 0. \tag{2.11}$$

Next we study analogously the action for a free massless antisymmetric tensor potential field, $A^{\mu\nu}\left(=-A^{\nu\mu}\right)$, in $D$ space-time dimensions.[12]

$$S_{(2)} = \kappa_D \int d^D x \left( \tfrac{1}{6} F^{\mu\nu\rho} F_{\mu\nu\rho} \right). \tag{2.12}$$

Here

$$F^{\mu\nu\rho} \equiv \partial^\mu A^{\nu\rho} + \partial^\nu A^{\rho\mu} + \partial^\rho A^{\mu\nu} \tag{2.13}$$

is the fully antisymmetric third rank tensor field strength obtained from the antisymmetric tensor potential. Finding an extremum of the action (2.12) under the variation $\delta A_{\nu\rho}$ leads to Euler-Lagrange equations which parallel (2.8),

$$\partial_\mu F^{\mu\nu\rho} = 0. \tag{2.14}$$

The highlight of the antisymmetric tensor field theoretic action is the local gauge invariance. Thus, if [13]

$$\delta A^{\mu\nu} = \partial^\mu \lambda^\nu - \partial^\nu \lambda^\mu, \tag{2.15}$$

then

$$\delta F^{\mu\nu\rho} = 0, \tag{2.16}$$

and

$$\delta S_{(2)} = 0. \tag{2.17}$$

Now consider adding analogous self-interaction terms to $S_{(1)}$ and $S_{(2)}$ respectively:

$$S_{m_1} = \kappa_D \int d^D x \left( -\tfrac{1}{2} m_1^2 A^\mu A_\mu \right), \tag{2.18}$$

and[14]

$$S_{m_2} = \kappa_D \int d^D x \left( \tfrac{1}{2} m_2^2 A^{\mu\nu} A_{\mu\nu} \right). \tag{2.19}$$

Simply added to their corresponding massless actions, these terms lead to Euler-Lagrange equations which describe a modified space-time field evolution.

To carry analogy to the traditional Higgs mechanism further and obtain an analogous action to (2.1) for $A^{\mu\nu}$ requires the definition of a combined field theoretic action that describes the propagation of, as well as the *interaction* between, a vector field and an antisymmetric tensor field. Also, let us only include a mass-like term for $A^{\mu\nu}$. An analogous candidate action is given by

$$S_{(1,2)} = \kappa_D \int d^D x \left( \tfrac{1}{6} F^{\mu\nu\rho} F_{\mu\nu\rho} - \tfrac{1}{4} F^{\mu\nu} F_{\mu\nu} + g A^{\mu\nu} F_{\mu\nu} - \tfrac{1}{2} m_2^2 A^{\mu\nu} A_{\mu\nu} \right). \tag{2.20}$$

Here $g$ is the coupling constant for the interaction between the vector field strength $F_{\mu\nu}$ and the antisymmetric tensor potential field $A^{\mu\nu}$. The generalized Higgs mechanism consists of postulating the continued invariance of the action $S_{(1,2)}$ under a combined gauge transformation of both potentials $A^{\mu}$ and $A^{\mu\nu}$. However, although we retain the form of the gauge transformation for $A^{\mu\nu}$ from (2.15), we modify that of $A^{\mu}$ from (2.9). The modification, in this case, is chosen to allow the vector to be "eaten" by the antisymmetric tensor $A^{\mu\nu}$ while simultaneously causing it have mass $m_2$, by way of satisfying a Klein-Gordon equation. Consider therefore the following set of joint transformations parameterized by $\lambda^{\mu}$, $\lambda$, and a constant[15] $b$

$$\delta A^{\mu\nu} = \partial^{\mu}\lambda^{\nu} - \partial^{\nu}\lambda^{\mu}, \tag{2.21}$$

$$\delta A^{\mu} = b\lambda^{\mu} + \partial^{\mu}\lambda. \tag{2.22}$$

If we choose

$$g = -m_2/\sqrt{2}, \tag{2.23}$$

and joint transformations with

$$b = m_2/\sqrt{2}, \tag{2.24}$$

we obtain

$$\delta S_{(1,2)} = 0. \tag{2.25}$$

Now, in general, the part of $\lambda^{\mu}$ that is expressible as a gradient does not contribute to (2.21) and may be absorbed into the gradient part of (2.22). Furthermore, the gradient part of $A^{\mu}$ does not contribute to the action because $S_{(1,2)}$ only depends on $A^{\mu}$ through $F^{\mu\nu}$.

Now let

$$B^{\mu\nu} \equiv A^{\mu\nu} + \frac{\sqrt{2}}{m_1} F^{\mu\nu}. \tag{2.26}$$

Thus, the action may be rewritten as

$$S_{(1,2)} = \kappa_D \int d^D x \left( \tfrac{1}{6} G^{\mu\nu\rho} G_{\mu\nu\rho} - \tfrac{1}{2} m_2^2 B^{\mu\nu} B_{\mu\nu} \right). \tag{2.27}$$

Where

$$G^{\mu\nu\rho} \equiv \partial^{\mu} B^{\nu\rho} + \partial^{\nu} B^{\rho\mu} + \partial^{\rho} B^{\mu\nu} = F^{\mu\nu\rho}. \tag{2.28}$$

From (2.27), the Euler-Lagrange equations of motion for $B^{\nu\rho}$ are

$$\partial_\mu \partial^\mu B^{\nu\rho} + m_2^2 B^{\nu\rho} = 0, \tag{2.29}$$

with Lorentz condition

$$\partial_\mu B^{\mu\nu} = 0. \tag{2.30}$$

Thus, an antisymmetric tensor potential field remains with a mass equal to $m_2$, is free of interactions, obeys the Lorentz condition, and the vector potential has been consumed.

The original discovery of this mechanism, as applied to vector and rank-2 antisymmetric tensor field, was accomplished in a hybrid theory involving fields mediating direct inter-string interactions between string bodies, including end-points.[16] We see here that the field theoretic process can stand alone provided that an appropriate partial action such as (2.20) is postulated. The result is also seen as completely analogous to the original scalar-vector Higgs mechanism for which the corresponding action is given in (2.1). We note in passing the feature of additional freedom due to the "nested" gauge transformation parameterized by $\lambda$.

Given the above developments coming from the study of $S_{(0,1)}$ (scalar/vector mechanism) and $S_{(1,2)}$ (vector/rank-2 antisymmetric tensor mechanism), we are led to consider actions containing still higher rank antisymmetric tensors and their corresponding Higgs mechanisms. Furthermore, the existence and properties of these actions, along with the two already studied, depend on the space-time dimensions in which they live. For example, if $D = 4$ we in principle also need to consider the mechanism in the action $S_{(2,3)}$. However, note that antisymmetric tensors (potentials or force fields) of rank higher than 4 cannot exist in a $D = 4$ space-time; thus, $S_{(2,3)}$ terminates the action sequence. In order to study $S_{(2,3)}$, we postulate the 3$^{\text{rd}}$ rank antisymmetric tensor potential field $A^{\mu\nu\rho}$ and its corresponding tensor force field $F^{\mu\nu\rho\sigma}$ and proceed as before. The set of actions in $D = 4$ space-time is designated as

$$\mathcal{A}_4 = \left\{ S_{(i-1,i)} \mid i = 1, 2, 3 \right\}. \tag{2.31}$$

These actions include field strengths defined in terms of their corresponding potentials

$$\begin{aligned}
F^\mu &\equiv \tfrac{1}{6} \varepsilon^{\mu\nu\rho\sigma} \varepsilon_{\nu\rho\sigma\delta} \partial^\delta A = \partial^\mu A, \\
F^{\mu\nu} &\equiv \tfrac{1}{2} \varepsilon^{\mu\nu\rho\sigma} \varepsilon_{\rho\sigma\gamma\delta} \partial^\delta A^\gamma = \partial^\mu A^\nu - \partial^\nu A^\mu, \\
F^{\mu\nu\rho} &\equiv \tfrac{1}{2} \varepsilon^{\mu\nu\rho\sigma} \varepsilon_{\sigma\beta\gamma\delta} \partial^\delta A^{\beta\gamma} = \partial^\mu A^{\nu\rho} + \partial^\nu A^{\rho\mu} + \partial^\rho A^{\mu\nu}, \\
F^{\mu\nu\rho\sigma} &\equiv \tfrac{1}{6} \varepsilon^{\mu\nu\rho\sigma} \varepsilon_{\alpha\beta\gamma\delta} \partial^\delta A^{\alpha\beta\gamma} = \tfrac{1}{6} \varepsilon^{\mu\nu\rho\sigma} \partial^\delta \tilde{A}_\delta,
\end{aligned} \tag{2.32}$$

with

$$\tilde{A}_\delta \equiv \varepsilon_{\alpha\beta\gamma\delta} A^{\alpha\beta\gamma}. \tag{2.33}$$

Here $\varepsilon^{\mu\nu\rho\sigma}$ is the Levi-Civita density for $D=4$, and $\varepsilon^{0123} = -\varepsilon_{0123} = 1$.

The set of antisymmetric fields and actions becomes more interesting when the number of space-time dimensions increases. Thus, we have in general

$$\begin{aligned}
\mathscr{P}_D &= \left\{ A^{\alpha_1\alpha_2\cdots\alpha_l} \,\middle|\, l = 0,1,\ldots,D-1 \right\}; \\
\mathscr{F}_D &= \left\{ F^{\mu_1\mu_2\cdots\mu_m} \equiv \tfrac{1}{(D-m)!(m-1)!} \varepsilon^{\mu_1\mu_2\cdots\mu_m\mu_{m+1}\cdots\mu_D} \varepsilon_{\mu_{m+1}\cdots\mu_D\alpha_1\alpha_2\cdots\alpha_{m-1}\beta} \partial^\beta A^{\alpha_1\alpha_2\cdots\alpha_{m-1}} \,\middle|\, m = 1,2,\ldots,D \right\}; \\
\mathscr{A}_D &= \Bigg\{ S_{(n-1,n)} = \kappa_D \int d^D x \, \tfrac{(-1)^n}{2} \bigg[ \tfrac{1}{n+1} F^{\mu_1\cdots\mu_n} F_{\mu_1\cdots\mu_n} - \tfrac{1}{n} F^{\mu_1\cdots\mu_{n-1}} F_{\mu_1\cdots\mu_{n-1}} \\
&\qquad - \tfrac{m_n}{\sqrt{n}} A^{\mu_1\cdots\mu_{n-1}} F_{\mu_1\cdots\mu_{n-1}} - \tfrac{1}{2} m_n^2 A^{\mu_1\cdots\mu_{n-1}} A_{\mu_1\cdots\mu_{n-1}} \bigg] \,\bigg|\, n = 1,2,3,\ldots,D-1 \Bigg\}.
\end{aligned} \tag{2.34}$$

We may define $D$ antisymmetric potentials in $\mathscr{P}_D$,[17] derive $D$ antisymmetric field strength tensors in $\mathscr{F}_D$, and construct $D-1$ actions in $\mathscr{A}_D$ that can support the analogous mechanisms. Each field strength tensor carries a gauge invariance, and each gauge invariance has its own nested invariance because of its antisymmetric form; $m_n$ is the mass value that each mechanism produces, which at this point may be different for each mechanism action. Notice that the interaction term contains a factor of $1/\sqrt{n}$, and the Levi-Civita densities are each shown explicitly with $D$ suffixes. Observe, finally, that as $D$ increases, the number of fields and their respective number of indices increases as well. Thus, additional interactions may be contemplated, and $\mathscr{A}_D$ may indeed be a subset of a more complete set of invariant actions displaying the Higgs mechanisms.

### III. CONCLUSION

The preceding developments have focused on the interesting physics of antisymmetric tensor field theories. The prototype rank-2 version, originally discovered as an interaction potential between the bodies of relativistic strings in $D=4$, was itself motivated by analogy with the Maxwell-vector (rank-1) interaction between point particles. We have further defined rank-$n$ antisymmetric tensors in $D$-dimensional space-times. However, the tensors require $n < D$ to produce a nontrivial theory. This relates directly to the association of each rank-$n$ antisymmetric tensor with a geometric object [$(n-1)$-brane], whose world-volume has dimension $n$. An object with $D$ dimensions is trivial since the corresponding field strength vanishes – there is no space for dynamics.

We also observed that each field strength Lorentz tensor remains invariant under a set of internal symmetry transformations of its underlying potentials. In the rank-1 case this is just ordinary

Maxwell gauge invariance, and we described the generalized notion of gauge invariance to the higher rank cases. It is intriguing that these gauge invariances nest rank-by-rank.

With the tensor field sequences, ranks, and the gauge invariances established, we found that combining any two sequential field theories, supplemented by a manifestly relativistically invariant coupling term containing the field strength of one and the potential field of the other, produced an action subject to a generalized Higgs mechanism. Clearly, the number of such actions depends on the postulated dimension of space-time. As expected, each action described a massive field theory of the higher rank antisymmetric field therein contained, along with a Lorentz condition. The interacting lower ranked field was consumed.

Going forward, in order to abstract further field theoretic behaviors, we would find it useful to add additional terms that maintain the generalized gauge invariance of the actions. Moreover, it would be interesting to construct invariant actions that contain three or more fields, even perhaps the entire sequence for a given space-time dimension. If some of the higher space-time dimensions were compactified, how would these actions behave, and would artifacts of these dimensions remain? Finally, what do these developments imply regarding the properties of submicroscopic matter and energy as might be measured in future experiments?

We thank Joshua Kalb and Robert Shaw for useful conversations and comments.

---

[1] E. C. G. Stueckelberg, Helv. Phys. Acta **30**, 209 (1957); D.G. Boulware and W. Gilbert, Phys. Rev. **126**, 1563 (1962); G. S. Guralnik, C. R. Hagen, and T. W. Kibble Phys. Rev. Lett. **13**, 585 (1964).

[2] This quantity is commonly denoted as the Kalb-Ramond field, cf. Barton Zwiebach, *A First Course in String Theory*, Cambridge University Press, Cambridge, UK, (2004).

[3] Michael Kalb and P. Ramond, Phys. Rev. D **9**, 2273, (1974) and references contained therein.

[4] In this context the fields can be considered as background. Cf. Soo-Jong Rey, Phys. Rev. D **40**, 3396 (1989).

[5] *Op cit*.

[6] Much as the set of local gauge transformations of the vector field contains global transformations as a subset.

[7] The Large Hadron Collider (LHC) was expected to go on-line in 2008 and begin this search in earnest, cf. Michael Dine, *Physics Today*, 33, (12/2007). Functional delays now have start-up after Spring 2009.

[8] The subscript (0,1) indicates that the action contains Lorentz scalar (0) and vector (1) degrees of freedom.

[9] In our conventions $1 = g_{00} = -g_{ii}, i = 1,\ldots, D-1$. Also, $\hbar = c = 1$. The constant $\kappa_D$ is included to give invariant actions the correct units in space-time of dimension $D$.

[10] Only terms that contribute directly to the mechanism are included.

[11] Note that a constant $c$ may be added to $\lambda$ without disturbing the gauge invariance.

[12] In this work potential fields are designated by the symbols $A_{...}$, while the derived field strengths are designated by the symbols $F_{...}$. All potentials have the same units. Again a constant $\kappa_D$ is included to give invariant actions the correct units in space-time of dimension $D$.

[13] Note that a gradient $\partial^\mu \lambda$ may be added to $\lambda^\mu$ without disturbing the gauge invariance.

[14] Note the opposite signs of the additional terms. This stems from the signs of the respective derivative terms.

[15] The analogy to (2.2) is evident.

[16] *Op cit.*

[17] If the suffix $m = 0$, we have the scalar potential case.